\documentclass[superscriptaddress,floatfix,aps,pre,twocolumn,notitlepage,shortbibliography,nofootinbib]{revtex4-2}

\usepackage[utf8]{inputenc}
\usepackage{amsmath}
\usepackage{amssymb}
\usepackage{refcount}
\usepackage{siunitx}
\usepackage{graphicx}
\usepackage{hyperref}
\usepackage{natbib}
 \usepackage[capitalise]{cleveref}
\usepackage{microtype}
\usepackage{bm}
\usepackage{lipsum}
\usepackage[english]{babel}
\usepackage{color}
\usepackage{graphicx}
\usepackage{xcolor}
\usepackage{braket}

\newcommand{\abs}[1]{\left| #1 \right|}

\newcommand{\bE}{\mathbf{E}}
\newcommand{\bB}{\mathbf{B}}

\newcommand{\bs}{\textbf{s}}

\begin{document}
\author{Haidar Al-Naseri}
\email{haidar.al-naseri@umu.se}
\affiliation{Department of Physics, Ume{\aa} University, SE--901 87 Ume{\aa}, Sweden}
\author{Gert Brodin}
\email{gert.brodin@umu.se}
\affiliation{Department of Physics, Ume{\aa} University, SE--901 87 Ume{\aa}, Sweden}
\title{Ponderomotive force due to the intrinsic spin for electrostatic waves in a magnetized plasma}
\begin{abstract}
We study the contribution from the electron spin to the ponderomotive force, using a quantum kinetic model including the spin-orbit correction. Specifically, we derive an analytical expression for the ponderomotive force, applicable for electrostatic waves propagating parallel to an external magnetic field. To evaluate the expression, we focus on the case of Langmuir waves and on the case of the spin-resonance wave mode, where the classical and spin contributions to the ponderomotive force are compared. Somewhat surprisingly, dependent on the parameter regime, we find that the spin contribution to the ponderomotive force may dominate for the Langmuir wave, whereas the classical contribution can dominate for the spin resonance mode. Naturally, this does not prevent the opposite case from being the more common one.  
   
\end{abstract}
\maketitle

\section{Introduction}
During the last decades, there has been an increasing number of works, see e.g. the reviews \cite{manfredi2019phase,shukla2011colloquium,melrose2020quantum,vladimirov2011description,Brodin2022Review} and references therein, studying quantum plasma physics. The motivation behind the works includes various applications, for example, quantum wells \cite{Quantumwells}, plasmonics \cite{Plasmonic} and spintronics \cite{Spintronics}, as well as astrophysics \cite{Astro1,Astro2}, strong field dynamics, and general theoretical interest. As a first rule of thumb, a quantum  description of plasmas is needed in the low-temperature high-density regime, as displayed  in temperature density plots made e.g. in Refs. \cite{shukla2011colloquium,manfredi2019phase}. However, it should be noted that quantum plasma behavior also can be introduced by a strong magnetic field such as in astrophysics (e.g. causing Landau quantization), and by strong laser fields inducing spin-polarization \cite{li2019ultrarelativistic,del2017spin}.   

The ponderomotive force is the main source behind broad classes of nonlinear plasma phenomena. Concrete examples include e.g. wake-field generation \cite{Wakefield1,Wakefield2}, soliton formation \cite{thejappa2018langmuir}, self-focusing \cite{Self-focusing}, and the subsequent nonlinear wave collapse \cite{kaw2017nonlinear}. Pioneering work regarding the classical expression for the ponderomotive force in a magnetized plasma was made by Karpman and Washimi \cite{Karpman} based on fluid theory, which was later generalized to include kinetic effects. The generalization of the ponderomotive force in magnetized plasma to include quantum effects, in particular, due to spin, has been done in Refs. \cite{brodin2010spin,stefan2011ponderomotive}. However, these works considered the effects due to non-relativistic spin dynamics. Moreover, in Ref. \cite{stefan2013linear}, the ponderomotive force due to semi-relativistic spin dynamics in unmagnetized plasma was calculated. 

In this work, we calculate the ponderomotive force due to semi-relativistic spin dynamic in magnetized plasmas. To be more specific, we consider electrostatic waves using the kinetic equation derived by Asenjo et al. \cite{Asenjo_2012}. Firstly, we study the linear electrostatic wave propagation in a magnetized plasma. This results in deriving the dispersion relation for the electrostatic waves. In addition to the common Langmuir mode, even for the case of immobile ions, it should be noted that in a magnetized plasma linearized theory allows for a new spin-dependent wave mode referred to as the spin resonance mode.  In section II C, we use perturbation theory based on linear calculations in order to calculate the ponderomotive force. Next, in section III, the general result is evaluated, comparing the magnitude of the classical and of spin-dependent contributions. This comparison is split into two parts, depending on whether the linear wave mode is a Langmuir wave or a spin resonance mode. Finally, in section IV, the results are summarized and the conclusions are drawn.    

\section{Basic equations and derivations}
In this section, we first present the basic quantum kinetic theory to be used throughout the manuscript. The theory is then used to investigate the linearized eigenmodes in a magnetized plasma in an electrostatic field geometry with the wave vector parallel to the external magnetic field. In the next sub-section, we perform nonlinear perturbation theory based on previous results, in order to deduce the ponderomotive force for electrostatic waves.  
\subsection{Basic equations}
Different quantum kinetic theories have been put forward in the literature, see e.g. the reviews given in \cite{Brodin2022Review}. In particular, two models that have been proven to be equivalent, based on the weakly relativistic limit of the Dirac Hamiltonian, have been derived in Ref. \cite{Asenjo_2012} and in Ref. \cite{hurst2017phase}. We will make use of the former formulation, based on a scalar distribution function, where the usual phase space is extended by a dependence on the independent spin-variable \cite{Asenjo_2012}.  Specifically, we will use the governing equation 
\begin{multline}
\label{Kinetic equation}
    \frac{\partial f}{\partial t}
    + \Big[\frac{\textbf{p}}{m} + \frac{\mu }{2mc} \bE \times (\bs +\nabla_s) \Big]\cdot \nabla_xf\\
    +q  \bigg(\bE 
    + \frac{1}{c}\Big[\frac{\textbf{p}}{m} + \frac{\mu }{2mc} \bE \times (\bs +\nabla_s) \Big]\times \bB
    \bigg)\cdot \nabla_pf\\
    +\frac{2\mu}{\hbar} \bs \times \Big(\bB- \frac{\textbf{p} \times \bE}{2mc}
    \Big)\cdot \nabla_sf\\
    +\mu\nabla_x \bigg[  (\bs + \nabla_s)\cdot\Big(\bB- \frac{\textbf{p} \times \bE}{2mc}
    \Big) \bigg]\cdot \nabla_p f=0,
\end{multline}
where $f(x,p,s,t)$ is the quasi-distribution function in phase-space, extended by the independent spin-variable $\bs$, defined to have unit length, $m$ is the electron mass, $\mu=\hbar q/2mc$ is the electron magnetic moment and $q=-e$ is the electron charge. 
This equation describes the dynamics of an ensemble of spin-1/2 particles in the Hartree approximation, i.e. the derivation applies mean-field theory neglecting correlations and exchange effects. 

While this model contains most dynamical effects related to the electron spin, such as the magnetic dipole force, spin precession, and the spin-orbit interaction, the evolution equation still neglects particle dispersive effects. This is a valid approximation in the regime of relatively long scale-lengths, fulfilling $\hbar^2\nabla_x^2\nabla_p^2\ll1$.  Note that we have also omitted the Darwin-term in the original kinetic equation derived by \cite{Asenjo_2012} since it is smaller than the other terms in the regime of consideration, with $\hbar^2 \nabla_x^2\ll m^2c^2$. 
Furthermore, since the model is semi-relativistic, the relation between $\textbf{v}$ and $\textbf{p}$ is non-trivial, reading
\begin{equation}
    {\bf v}=\frac{\textbf{p}}{m} + \frac{3\mu}{2mc}\bE \times \bs
\end{equation}
This relation is important when the sources in Maxwell's equation is computed. The relations needed to close the system are as follows
\begin{align}
    \nabla \cdot \bE&= 4\pi \rho\\
   \nabla \times \bB &= \frac{1}{c}\frac{\partial \bE}{\partial t} +\frac{4\pi}{c} \textbf{J} \label{Ampers_Law} , 
\end{align}
where $\rho$ and $\textbf{J}$ are the charge and current density 
\begin{align}
    \rho&= \rho_f+ \nabla \cdot \textbf{P}\\
    \textbf{J}&= \textbf{J}_f+ \nabla \times \textbf{M}+\frac{\partial \textbf{P}}{\partial t} \label{Total_current},
\end{align}
where 
\begin{align}
    \rho_f&= q\int d \Omega f\\
      \textbf{P}&= -3\mu \int d\Omega\frac{\bs \times \textbf{p}}{2mc} f\label{Polarization}\\
      \textbf{J}_f&=q\int d\Omega \Big[\frac{\textbf{p}}{m} + \frac{3\mu}{2mc}\bE \times \bs \Big]f\label{Free_current}\\
      \textbf{M}&= 3\mu \int d\Omega\, \bs f
\end{align}
are where the expressions represent the free charge density, the polarization, the free current density, and the magnetization, respectively. Here, we have used $d\Omega=d^3pd^3xd^2s$.
In this work, we express the momentum $\textbf{p}$ in cylindrical coordinates $(p_{\bot},\phi_p,p_z)$   while for spin $\bs$, we use spherical coordinates $(\phi_s,\theta_s)$.  

Before we proceed with the analysis, let us point out that the closely related model derived by \cite{hurst2017phase} does not use spin as an independent variable, but instead has a classical type of (scalar) distribution function for the charge density, and a vector-valued distribution function for the magnetization. The relation between these two models has been described in some detail in Refs. \cite{Brodin2022Review,manfredi2019phase}. We stress that although the models are technically different, they have been shown to be formally equivalent.      

\subsection{Linear theory}

As a prerequisite to computing the ponderomotive force, we first study the linearized theory. Specifically, we concentrate on electrostatic waves propagating parallel to an external magnetic field.
Thus we first divide the distribution function into $f(x,p,s,t)=f_0(p^2,\theta_s)+f_1(x,p,s,t)$, where $f_0$ is the background distribution function which is homogeneous (see e.g. Ref. \cite{Zamanian_2010, Brodin2022Review} for a discussion of possible background functions) and $f_1$ is the perturbed distribution function. The dependence $f_0(p^2)$  assures that the momentum dependence of the background is isotropic. 
For our case of electrostatic waves propagating parallel to an external magnetic fields we have
\begin{align*}
\textbf{E}&=E \,\hat{\textbf{z}}\\ \textbf{B}&=B_0\,\hat{\textbf{z}}\\ \textbf{k}&=k\,\hat{\textbf{z}}. 
\end{align*}
To proceed, we use spherical coordinates in spin space $\phi_s,\theta_s$, with the length $\abs{s}=1$. Thus the Cartesian components are written ${\bf s}=(\cos\phi_s \sin\theta_s,\sin\phi_s \sin\theta_s,cos\theta_s)$. Next, 
we linearize \cref{Kinetic equation} and expand $f_1$ using the following ansatz  
\begin{equation}
    f_1= \frac{1}{2\pi} \sum_{n,n'=-\infty}^{\infty}g_{n,n'}e^{i n\varphi_p}e^{in'\varphi_s} e^{i(kz-\omega t)}
\end{equation}
Applying this ansatz to the linearized version of \cref{Kinetic equation}, after some algebra we find an explicit expression of $f_1$ in terms of the unperturbed function $f_0$ of the form
\begin{equation}
\label{Linear_Solution}
    f_1= A+ B_{+}e^{i(\varphi_p-\varphi_s)} + B_{-}e^{-i(\varphi_p-\varphi_s)}
\end{equation}
where 
\begin{align}
A&= -\frac{iqE}{\omega -kp_z/m} \frac{ \partial f_0}{\partial p_z}
  \\
    B_{\pm}&= -i \frac{q\mu B_0 E /4mc}{\omega -kp_z/m \mp \Delta\omega_{ce} } \Big(\sin\theta_s+\cos\theta_s
    \frac{\partial}{\partial_{\theta_s}} \Big) \frac{ \partial f_0}{\partial p_{\bot}}\notag\\
    &\pm i \frac{k\mu E p_{\bot} /4mc}{\omega -kp_z/m \mp \Delta\omega_{ce} } \Big(\sin\theta_s+\cos\theta_s
    \frac{\partial}{\partial_{\theta_s}} \Big)\frac{ \partial f_0}{\partial p_z}\notag\\
    &+i \frac{\mu E p_{\bot}/2\hbar mc}{\omega -kp_z/m \mp \Delta\omega_{ce} }
    \frac{ \partial f_0}{\partial \theta_s }.
\end{align}
Here, $\Delta \omega_{ce}=\omega_{cg}- \omega_{ce}$, $\omega_{ce}=qB/m$ is the cyclotron frequency and $\omega_{cg}=(g/2)\omega_{ce}$ is the spin precession frequency and $g\approx 2.002318$ is the electron g-factor. Note that in the classical ( $\hbar \longrightarrow 0$) limit, we get $B_{\pm}=0$ and we have the standard classical expression for electrostatic Langmuir waves.

Next we calculate the dispersion relation by using Ampérs law \cref{Ampers_Law}, where the total current ${\bf J}$ is given by \cref{Total_current}.
In the integration process when calculating the currents, we expand the denominators in \cref{Linear_Solution} to the first non-vanishing order of $p_z$, as is appropriate for a low or modest temperature. This condition is also necessary to avoid strong wave-particle interaction leading to appreciable wave-damping.
Moreover, we use the following expression of the background distribution $f_0$
\begin{equation}
\label{Background_Dis}
f_0(p^2,\theta_s)=\sum_{\pm}(1\pm \cos\theta_s)f_{0\pm}(p^2),
\end{equation}
where $f_{0\pm}(p^2)$ is the unperturbed distribution function for the particles in spin up/down state. Thus we have  $\int d\Omega f_{0\pm}(p^2)=n_{0\pm}$, where $n_{0\pm}$ is the number density for spin up/down state. To carry the momentum integration, we need to specify the background distribution function $f_{0\pm}$. For a non-degenerate plasma, where the Fermi temperature is well below the thermodynamic temperature, the appropriate distribution function is the Maxwell-Boltzmann distribution function with a spin-dependent part \cite{Zamanian_2010}
\begin{equation}
    f_{0\pm}= \frac{1}{N_m}\,e^{-p^2/m^2v_{th}^2}\,e^{\pm \mu B_0/K_B T} ,
\end{equation}
where $K_B$ is the Boltzmann constant, $T$ is the temperature, the thermal velocity $v_{th}$ fulfills  $mv_{th}^2/2=k_B T$, and $N_m=8m^3v_{th}^3\pi ^{5/2} \cosh{(\mu B_0/K_BT)} $ is the normalization factor.
After carrying out the spin- and momentum-integration, we obtain the dispersion relation
\begin{widetext}

\begin{multline}
\label{Dispersion_rel}
    \omega^2 \Bigg(
    1
    + \frac{\hbar^2\omega_p^2 \Delta\omega_{ce}}{8m^2c^4}
    \Bigg[
    \frac{\omega_{ce}}{\omega^2-\Delta\omega_{ce}^2}
    +\frac{k^2v_{th}^2/2(3\omega^2\omega_{ce}+\Delta\omega_{ce}^2\omega_{ce}
    )}{(\omega+\Delta\omega_{ce})^3(\omega-\Delta\omega_{ce})^3}
    \Bigg]
    + \frac{\omega m vt_{th}^2}{\hbar \Delta\omega_{ce}(\omega ^2-\Delta \omega_{ce}^2)}\tanh{\frac{\mu B_0}{K_BT}}\\
    +\frac{k^2v_{th}^2 \omega}{(\omega +\Delta\omega_{ce})^2(\omega -\Delta\omega_{ce})^2}
    \Bigg)
    =\omega_p^2\Big(
    1
    +\frac{3}{2}\frac{k^2v_{th}^2}{\omega^2}
    \Big)
\end{multline}
\end{widetext}
where we have used
\begin{equation}
    \omega_p^2= \frac{q^2}{m} \sum_{\nu} \int d\Omega f_{0\nu},
\end{equation}
as the definition of the plasma frequency.
Taking the classical limit by letting $\hbar \rightarrow 0$ in \cref{Dispersion_rel}, most terms disappear and we get the classical Langmuir dispersion relation. While the coefficients in front of the spin-dependent terms are usually small (unless we have very high densities and/or magnetic field strengths), nevertheless the quantum terms can be important for wave-frequencies close to $\Delta\omega_{ce}$. The effects of spin-resonances, i.e. frequencies fulfilling $\omega\approx \Delta\omega_{ce}$ will be explored below. 

\subsection{The ponderomotive Force}

The aim of this sub-section is to generalize the linearized treatment to the weakly nonlinear regime, in order to deduce the ponderomotive force for electrostatic waves. For this purpose, we use the following ansatz
\begin{multline}
      f(x,p,s,t)=f_0(p^2,\theta_s)+ f_{lf}(z,t,p,\theta_s) 
    \\
    + \frac{1}{2} \Big[\Tilde{ f}_1(z,t,p,s)e^{ikz-i\omega t} +\Tilde{f}_1^{*}(z,t,p,s)e^{-ikz-i\omega t}
    \Big].
\end{multline}
to  calculate the weakly nonlinear low-frequency response to electrostatic waves. Here $f_{lf}$ is the low-frequency response due to quadratic nonlinearities, $\Tilde{f}_1$ represents the slowly varying high-frequency wave and the star denotes the complex conjugate. As usual, "slowly varying" means that the amplitude derivatives are small compared to the rapidly oscillating scale at $(\omega,k)$. Using this ansatz in \cref{Kinetic equation}, keeping up to quadratically nonlinear terms, and averaging to isolate the low-frequency scale, we obtain

\begin{multline}
    \label{LF}
    \Big[\partial_t + \frac{p_z}{m}\partial_z\Big]f_{lf}=
    -qE_{lf}\frac{\partial f_0}{\partial p_z}
    -\frac{\mu}{2mc} \Big[ \Tilde{\bE} \times (\bs +\nabla_s)\Big]\cdot \nabla_x \Tilde{f}_1^{*}\\
   - \frac{q\Tilde{E}}{4}\frac{\partial \Tilde{f}_1^{*}}{\partial p_z}
   -\frac{q\mu}{8mc} \Big( \big[ \Tilde{\bE} \times (\bs +\nabla_s)\big]\times \bB_0 \Big) \cdot \nabla_p\Tilde{f}_1^{*}\\
+\frac{\mu}{4\hbar mc} \Big[\bs \times (\textbf{p}\times \Tilde{\bE})  \Big]\nabla_s \Tilde{f}_1^{*}
+\frac{\mu}{8mc} \nabla \Big[(\bs+\nabla_s)  \cdot(\textbf{p}\times \Tilde{\bE})  \Big]\cdot \nabla_p\Tilde{f}_1^{*}\\
   + c.c.
\end{multline} 
The high frequency response $\Tilde{f}_1$ is obtained by making the substitution 
\begin{align*}
    \omega \rightarrow \omega +i\partial_t\\
    k \rightarrow k - i\partial_z
\end{align*}
in the linear solution of $f_1$ in \cref{Linear_Solution}, where $i\partial_t$ and $i\partial_z$ can be treated as small perturbations due to the slowly varying amplitudes. Now having an implicit expression for $f_{lf}$, we will calculate the total low-frequency current
\begin{equation}
    J_{lf}= J_{lf}^f+ J_{lf}^p,
\end{equation}
where
\begin{align}
\label{Current_low}
    J_{lf}^p&=-3\mu\partial_t \int d\Omega\frac{p_{\bot}}{2mc}\sin\theta_s \Big(\cos\varphi_s\sin\varphi_p \notag \\ &-\sin\varphi_s\cos\varphi_p\Big)f_{lf} \notag\\ 
    J_{lf}^f&= q\int d\Omega \frac{p_z}{m}f_{lf} 
\end{align}
are the free and polarization low-frequency current respectively. Note that the low frequency free current looks simpler than the expression in \cref{Free_current} since the current is directed along $\hat{z}$. 
Now we want to use the expression of $f_{lf}$ in \cref{LF} to calculate the current \cref{Current_low}. But since \cref{LF} does not provide an explicit expression of $f_{lf}$, we need to make some further calculations. We note that for low-frequency free current in \cref{Current_low}, we have the following relation
\begin{equation}
\label{Free_current_calc}
    \partial_t J_{lf}^{f} + q\int d\Omega \frac{p_z^2}{m^2}\frac{\partial f_{lf}}{\partial z} =
    q\int d\Omega \frac{p_z}{m}
    \Big[\partial_t + \frac{p_z}{m} \partial_z  \Big]  f_{lf}.
\end{equation}
The term in the square brackets in \cref{Free_current_calc} is the same as in the right hand side of \cref{LF}. However we have the integral in the left hand side of \cref{Free_current_calc} that we need to deal with. Due to the proportionality of $p_z^2$, this term is small in the low temperature limit and we will use this for a perturbative calculation in the next step.  
Taking the time-derivative of \cref{Free_current_calc}, we get 
\begin{multline}
\label{Free_current_lf}
    \partial_{t}^2 J^{f}_{lf}\approx
    q\partial_t\int d\Omega \frac{p_z}{m}
    \Big[\partial_t + \frac{p_z}{m} \partial_z  \Big]  f_{lf}\\
    -q\partial_z\int d\Omega \frac{p_z^2}{m^2}
    \Big[\partial_t + \frac{p_z}{m} \partial_z  \Big]  f_{lf}
\end{multline}
Note that we added $p_z^3/m^3 \partial^2_{z} f_{lf}$ in the last term. This term turns to be a higher order thermal correction to the rest of the terms, but we added it in order to use the implicit expression of $f_{lf}$ in \cref{LF}.
Doing the same procedure for the polarization current, we get
\begin{widetext}

\begin{multline}
\label{Polarization_Current_lf}
    \partial_{t}^2 J_{lf}^p \approx
    -\frac{3\mu}{2mc}\partial_t^2 \int d\Omega p_{\bot}\sin\theta_s (\cos\varphi_s\sin\varphi_p - \sin\varphi_s\cos\varphi_p)\Big[\partial_t + \frac{p_z}{m} \partial_z  \Big]f_{lf}\\
    +\frac{3\mu}{2mc}\partial_t \int d\Omega p_{\bot}\sin\theta_s (\cos\varphi_s\sin\varphi_p - \sin\varphi_s\cos\varphi_p)\frac{p_z}{m}\partial_z
    \Big[\partial_t + \frac{p_z}{m} \partial_z  \Big]f_{lf}
\end{multline}
\end{widetext}
Now we can calculate the integrals in \cref{Free_current_lf} and \cref{Polarization_Current_lf}. In doing that, we use the \cref{Background_Dis} for $f_0$ . Calculating the $\phi_s$, $\phi_p$ and $\theta_s$-integrals, we get
\begin{widetext}

\begin{multline}
\label{Free_Current_Int}
    \partial_t^2J_{lf}^f= -2(2\pi)^2q^2\sum_{\nu}\int p_{\bot}dp_{\bot}dp_z
    \Big[\frac{p_z}{m}\partial_t- \frac{p_z^2}{m^2}\partial_z  \Big]
    \Bigg[  
    E_{lf}\frac{\partial f_{0\nu}}{\partial p_z}\\
    + i \abs{E}^2 \frac{\partial}{\partial p_z}
    \bigg(
    \frac{1}{\omega - k\frac{p_z}{m} - i(\partial_t+ \frac{p_z}{m}\partial_z)}
    -    \frac{1}{\omega - k\frac{p_z}{m} + i(\partial_t+ \frac{p_z}{m} \partial_z)}
    \bigg)
    \frac{\partial f_{0\nu}}{\partial p_z}\\
    +i \abs{E}^2 \frac{\mu B_0}{32mc}\frac{\partial}{\partial p_{\bot}} \sum_{\pm} 
    \bigg(
    \frac{q\mu B_0 \partial_{p_{\bot}}f_{0\nu} \mp (k-i\partial_z)\mu p_{\bot}\partial_{p_z}f_{0\nu} + 2\nu \mu p_{\bot}/\hbar f_{0\nu}  }{\omega - k\frac{p_z}{m} - i(\partial_t+ \frac{p_z}{m}\partial_z)\mp \Delta\omega_{ce}}\\
    -    \frac{q\mu B_0 \partial_{p_{\bot}}f_{0\nu} \mp (k+i\partial_z)\mu p_{\bot}\partial_{p_z}f_{0\nu} + 2\nu \mu p_{\bot}/\hbar f_{0\nu} }{\omega - k\frac{p_z}{m} + i(\partial_t+ \frac{p_z}{m} \partial_z)\mp\Delta\omega_{ce}}
    \bigg)
    \Bigg]
\end{multline}
and
\begin{multline}
\label{Polarization_Current_Int}
    \partial_t^2 J_{lf}^p=  
\frac{   (2\pi)^2q^2\mu^2 \abs{E}^2 \partial_t  }{4m^2c^2}\sum_{\nu, \pm}\int p_{\bot}^2dp_{\bot}dp_z
   \Big[\partial_t-  \frac{p_z}{m}\partial_z \Big]
   \frac{\partial}{\partial p_z}\\
   \bigg[
   \frac{\pm B_0 \partial_{p_{\bot}}f_{0\nu} - (k-i\partial_z)p_{\bot}/q \partial_{p_z}f_{0\nu} \pm 2\nu p_{\bot}/\hbar q f_{0\nu} }{\omega - k\frac{p_z}{m} - i(\partial_t+ \frac{p_z}{m} \partial_z)\mp\Delta\omega_{ce}}\\
   +  \frac{\pm B_0 \partial_{p_{\bot}}f_{0\nu} - (k+i\partial_z)p_{\bot}/q \partial_{p_z}f_{0\nu} 
   \pm 2\nu p_{\bot}/\hbar q f_{0\nu}}{\omega - k\frac{p_z}{m} + i(\partial_t+ \frac{p_z}{m} \partial_z)\mp\Delta\omega_{ce}}
   \bigg]
\end{multline}
\end{widetext}
Expanding the denominators in \cref{Free_Current_Int} and \cref{Polarization_Current_Int} to lowest non-vanishing order of $p_z$, this is consistent with the approximation made in \cref{Free_current_lf} and \cref{Polarization_Current_lf}.
Then, we integrate over $p_z$ and $p_{\bot}$ and use Ampere's law

\begin{multline}
\label{Ponder_Force}
    \Big(\frac{\partial ^2}{\partial_t^2} + \omega_p^2  \Big)E_{lf}= -\frac{2 q\omega_p^2}{m\omega^2}
    \bigg[1- \frac{7\mu B_0\, \hbar \omega^2\Delta \omega_{ce} }{64 m^2c^3  (\omega^2-\Delta\omega_{ce}^2) }    \bigg] \frac{\partial \abs{E}^2} {\partial z}\\
    + \frac{\mu B_0 \hbar k q  }{16 m^3 c^3  }\,
    \frac{\omega \Delta
    \omega_{ce}\omega_p^2 }{ (\omega+\Delta\omega_{ce})^2
    (\omega-\Delta\omega_{ce})^2
    } 
    \frac{\partial \abs{E}^2}{\partial t},
\end{multline}

Taking the classical limit $\hbar \rightarrow 0$, we get
\begin{equation}
    \Big(\frac{\partial ^2}{\partial_t^2} + \omega_p^2  \Big)E_{lf}= -\frac{2 q\omega_p^2}{m\omega^2}
 \frac{\partial \abs{E}^2} {\partial z}
 \equiv \frac{qn_0}{\epsilon_0}f_p
\end{equation}
where $f_p$ is defined by the second equality, such that $f_p$ gives us the classical ponderomotive force. Due to the velocity perturbation being parallel to the external magnetic field, the unperturbed magnetic field does not influence the result in the classical case. However, quantum mechanically, due to the spin-orbit interaction, there is a contribution that modifies the classical ponderomotive force rather significantly, as seen by \cref{Ponder_Force}. In particular, in addition to a term proportional to the spatial intensity gradient, we get a term proportional to the temporal intensity gradient. Even more importantly, the quantum mechanical terms contain spin-resonances, that will be investigated in the next section.     

\section{Comparison of classical and non-classical contributions to the Ponderomotive force}

The purpose of this section is to illustrate the importance of the spin
contributions in \cref{Ponder_Force} by comparing the new terms to the
classical contribution. However, the relative magnitude of the spin terms
depend to a considerable degree on the linear wave properties of the
electrostatic pulse, as described by the dispersion relation \cref%
{Dispersion_rel}. To simplify the expression of the ponderomotive force in %
\cref{Ponder_Force}, we use that, to the lowest order approximation, the
pulse is stationary in a frame moving with the group velocity $v_{g}=\frac{%
\partial \omega }{\partial k}$, such that the approximation 
\begin{equation}
\frac{\partial \left\vert E\right\vert ^{2}}{\partial t}\simeq -v_{g}\frac{%
\partial \left\vert E\right\vert ^{2}}{\partial z}  \label{Derivative-approx}
\end{equation}%
can be applied to compare the magnitude of the terms in Eq. (\ref%
{Ponder_Force}). Thus, as a prerequisite to studying the ponderomotive
fore, we need to analyze the linear dispersion relation to deduce the group
velocity. While the general behavior of \cref{Dispersion_rel} can be
complicated, our analysis is simplified by the fact that for most naturally
occurring plasmas, we can treat $\hbar \omega _{p}/mc^{2}$ and $\hbar \omega
_{c}/mc^{2}$ as small parameters. For this case, that we focus on below, the
solutions of \cref{Dispersion_rel} separates into two modes. One mode
resembling the classical Langmuir mode to a good approximation, and another
mode with a frequency close to the spin resonance, approximately given by $%
\omega \simeq $ $\Delta \omega _{ce}$. We will simply refer to these modes
as the Langmuir mode and the spin resonance mode, respectively. Below we
compare the relative contribution of the classical and non-classical terms
in Eq. (\ref{Ponder_Force}) for the Langmuir mode and for the spin resonance
mode.

\subsection{The Langmuir mode}

For a case where the linear dispersion is approximately classical, the
frequency $\omega $ cannot be too close to the spin resonance. If the spin
resonance is avoided, however, the magnitude of the quantum terms in Eq. (%
\ref{Ponder_Force}) will also be somewhat limited. One would expect,
perhaps, that the condition for neglecting the spin contribution to \cref%
{Ponder_Force} would be the same as for dropping the spin contribution in %
\cref{Dispersion_rel}. As it turns out, however, this is not quite true. To
the contrary, it is possible to have a situation where the linear dispersion
relation is approximately classical, although the spin terms dominate the
expression for the ponderomotive force. This require an intermediate regime,
where the wave frequency is fairly close to the spin resonance, in order for
the spin contributions of Eq. (\ref{Ponder_Force}) to be magnified. Still, the wave frequency must be sufficiently far from the spin resonance, in
order not to invalidate the classical approximation of \cref{Dispersion_rel}%
. Firstly, we analyze the linear dispersion relation \cref{Dispersion_rel}, comparing the classical terms with the dominant spin
term. For the classical Langmuir dispersion relation to hold approximately,
we must have the strong inequality 
\begin{equation}
\frac{\hbar ^{2}\omega _{p}^{2}}{8m^{2}c^{4}}\frac{\omega _{ce}}{\Tilde{%
\omega}}\ll 1  \label{Langmuir inequality}
\end{equation}%
fulfilled, where $\Tilde{\omega}\equiv \omega -\Delta \omega _{ce}$. Assuming this
to hold, we can neglect all of the spin-terms in the dispersion relation %
\cref{Dispersion_rel}.

\begin{figure}[h!]
    \includegraphics [width=9 cm, height= 10 cm]{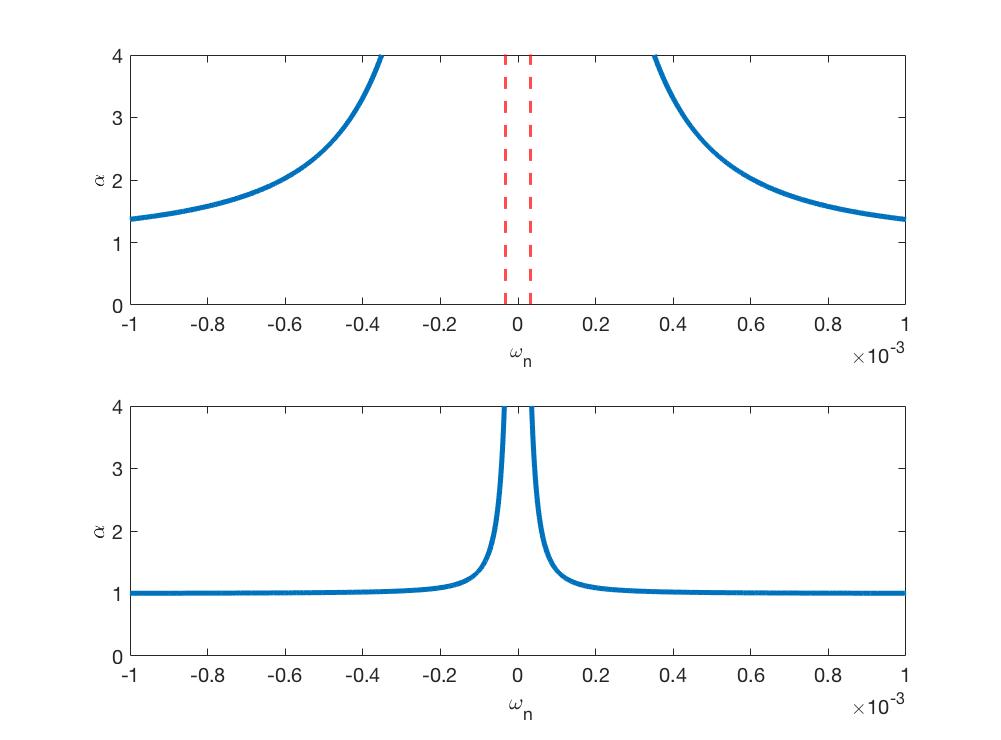}
    \caption{The fraction of the total and classical ponderomotive force $\alpha$ is plotted versus the normalized frequency difference $\omega_n= (\omega^2-\Delta \omega_{ce}^2)/ \Delta \omega_{ce}^2$ for $K_BT/mc^2=0.1$, $\omega_p/\Delta \omega_{ce}=0.7$ and $\mu B_0/mc^2= (0.1,0.01)$ for the upper and lower panels respectively.}
    \label{Pondclas}
\end{figure}

\begin{figure}[h!]
    \includegraphics [width=9cm,height=8cm]{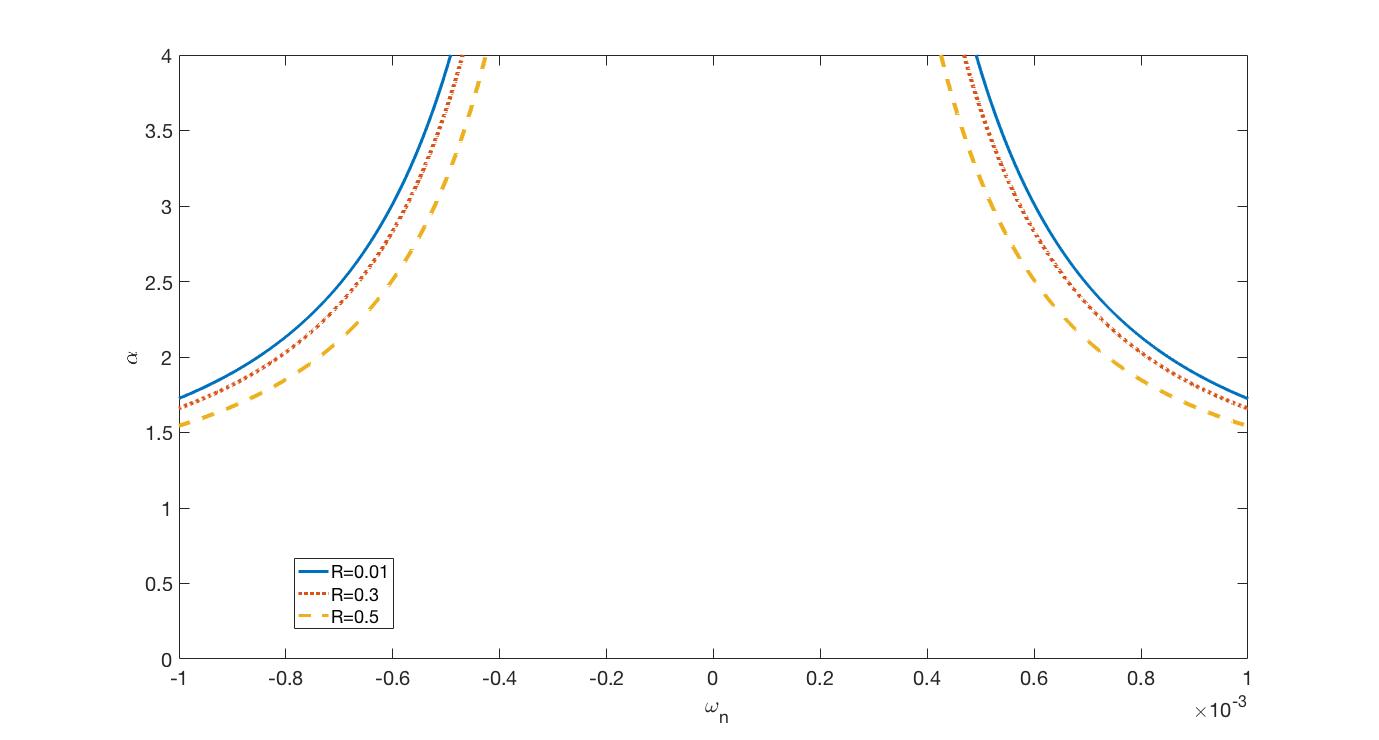}
    \caption{The fraction of the total and classical ponderomotive force $\alpha$ is plotted versus the normalized frequency difference $\omega_n= (\omega^2-\Delta \omega_{ce}^2)\Delta \omega_{ce}^2$ for different values of $R=\omega_p/\Delta\omega_{ce}$. }
    \label{Pondclas2}
\end{figure}
\begin{figure*}[]
    \includegraphics[ width=15 cm,height=9cm] {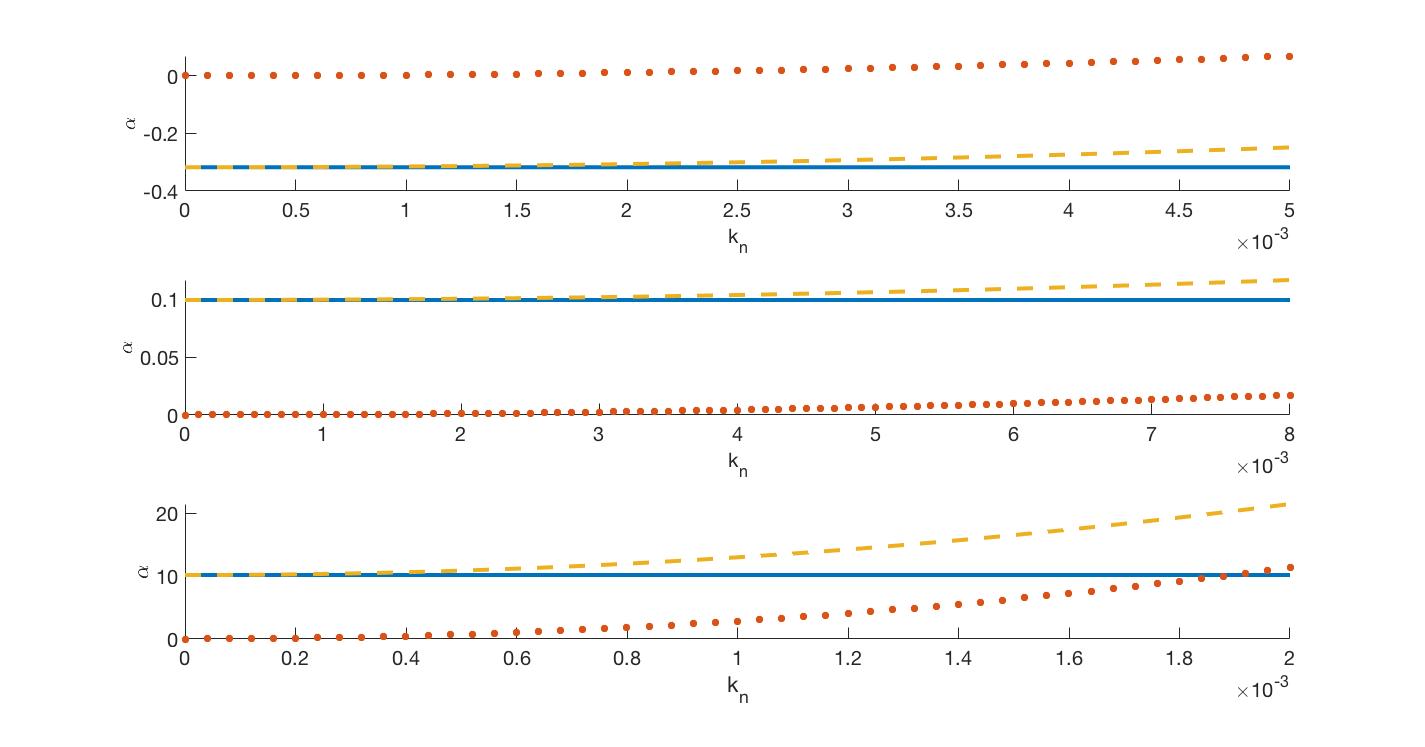}
    \caption{The fraction of the spin and classical ponderomotive force $\alpha$ is plotted versus the normalized wavelength $k_n= kv_{th}\Delta \omega_{ce}$. The solid curve is the first spin-term, the star-curve is the second spin term and the dashed curve is the total spin-force. In the first panel we have $B_n=0.1,R=2$, in the second we have $B_n=0.1,R=0.9$ and in the third $B_n=0.1, R=0.1$.}
    \label{fig3}
\end{figure*}
\begin{figure}[]
    \includegraphics [width=9 cm, height=9cm]{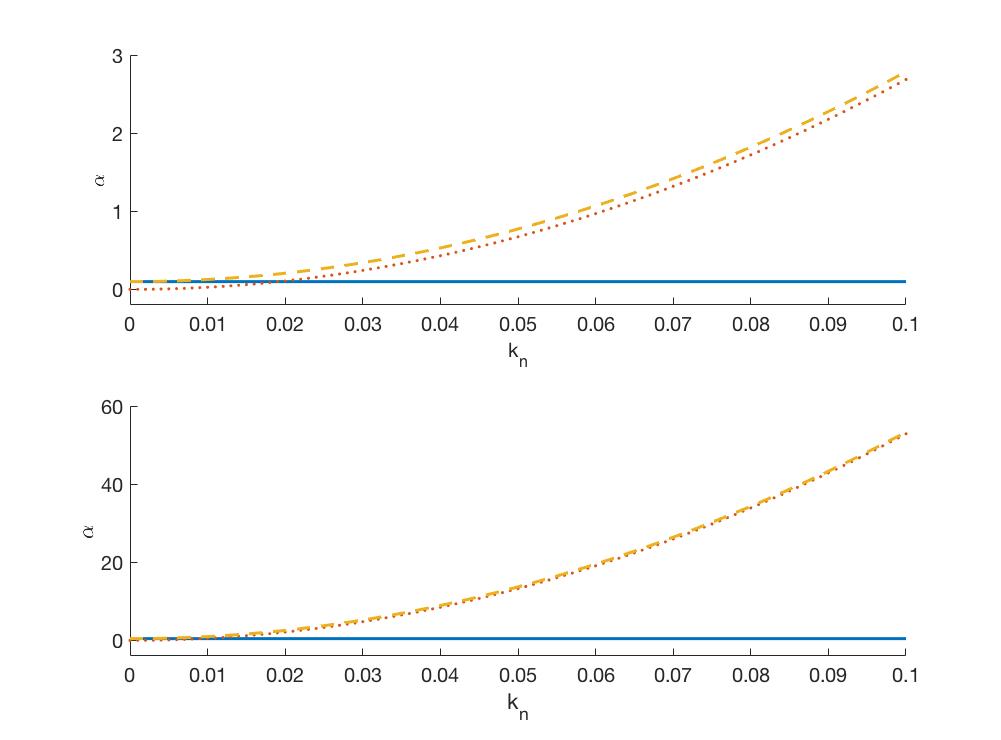}
    \caption{The fraction of the spin and classical ponderomotive force $\alpha$ is plotted versus the normalized wavelength $k_n= kv_{th}\Delta \omega_{ce}$. The solid curve is the first spin-term, the star-curve is the second spin term and the dashed curve is the total spin-force. In the first panel we have $B_n=0.1, R=0.9$ and in the second we have $B_n=0.1, R=0.7$.}
    \label{fig4}
\end{figure}

However, although we cannot be too close to the spin resonance (as implied
by \cref{Langmuir inequality}, we cannot be too far from the resonance, as
otherwise the spin terms will not be significant in Eq. (\ref{Ponder_Force}%
). In practice, for the Langmuir wave mode (and with $\hbar \omega
_{p}/mc^{2}\ll 1$ and $\hbar \omega _{c}/mc^{2}\ll 1$), spin terms are
significant only for a rather small wave number spectrum narrowly centered
around $k\simeq k_{c}$, where $k_{c}$ is the critical wave number where the
classical Lamgmuir dispersion coincides with the spin resonance frequency.
Thus, we will here be concerned with wave numbers $k\simeq k_{c}$, where $%
k_{c}$ fulfills: 
\begin{equation}
\omega ^{2}=\omega _{p}^{2}+\frac{3}{2}k_{c}^{2}v_{th}^{2}\approx \Delta
\omega _{ce}^{2}.
\end{equation}%
Evaluating the ponderomotive expression \cref{Ponder_Force} in a narrow
wave number spectrum centered around $k=k_{c}$ (such that $\omega $,
approximately given by the classical Langmuir dispersion relation is
centered around $\Delta \omega _{ce}$), we evaluate temporal derivatives
according to Eq. (\ref{Derivative-approx}). More specifically, in \cref{Pondclas}, we
plot the ratio of the total ponderomotive force and its classical
contribution only, for a narrow frequency spectrum surrounding the spin
resonance. We assume Eq. (\ref{Langmuir inequality}) to be fulfilled, such
that the classical Langmuir dispersion relation can be used to evaluate the
group velocity. Moreover, the small deviation of $k$ from $k_{c}$ has been
neglected in the plot. While the region where Eq. (\ref{Langmuir inequality}) 
is violated from the plot must be discarded (the region inside the two dashed 
vertical lines shown in the first panel of \cref{Pondclas}), we note that the spin terms
of Eq. (\ref{Ponder_Force}) magnifies the ponderomotive force in a wider
region frequency region than the one that must be excluded. In other words,
there is a narrow frequency band where the linear wave properties are
classical to a good approximation, but where the nonlinear properties need
to be evaluated with the spin terms included. In the second panel of \cref{Pondclas}, we see a similar plot, 
but for a somewhat weaker magnetic field (normalized magnetic 
field $B_n=\mu B_0/mc^2=0.01$), in which case the resonance region becomes slightly more narrow. 
The narrowing applies in a much higher degree to the validity condition. Thus, in the second panel of \cref{Pondclas}, 
the region violating the inequality \cref{Langmuir inequality} is too narrow to be displayed. 
To be concrete, if the two vertical lines were given as in \cref{Pondclas}, but for the new parameter values, 
the vertical lines would be centered too close to the precise resonance at $\omega_n=0$ to be separable in the given 
resolution. Finally, to show the role of a varying density, three curves for different values of 
$R\equiv \omega_p/\Delta \omega_{ce}$ are shown in \cref{Pondclas2}. We  can see that the curve with the higher value of 
$R$ has the most narrow resonance. 

Finally, before we turn our
attention to the spin resonance mode, we note that for the Langmuir mode,
the second spin term (proportional to $\partial \left\vert E\right\vert
^{2}/\partial t$) in \cref{Ponder_Force} always dominate over the first
(proportional to $\partial \left\vert E\right\vert ^{2}/\partial z$), since
the resonance is of a higher order for the second term.

\subsection{The Spin resonance mode}

Next, we will consider the complimentary frequency regime close to the spin
resonance where \cref{Langmuir inequality} is violated. Specifically,
we focus on the long wavelength regime $\left\vert kv_{th}/(\omega -\Delta
\omega _{ce})\right\vert \ll 1$, in order to avoid strong Landau damping of
the mode. Since we are focusing on the spin resonance mode, we can let $%
\omega =\Delta \omega _{ce}$ when evaluating the ponderomotive force terms \cref{Ponder_Force},
except in the denominators, where, obviously, a more accurate expression
must be used. As a prerequisite for further analysis, we calculate the
frequency for the mode at $k=0$, which will deviate slightly from $\Delta
\omega _{ce}.$ Approximating the denominators of $1/(\omega ^{2}-\Delta
\omega _{ce}^{2})$ of Eq. (\ref{Dispersion_rel}) as $1/[2(\omega -\Delta
\omega _{ce})\Delta \omega _{ce}]$, we compute the frequency for $k=0$ for
the spin resonance mode as%
\begin{equation}
\omega =\Delta \omega _{ce}\left( 1+\delta \right)  \label{Spin-res-mod}
\end{equation}
where%
\begin{equation}
\delta =\frac{\hbar ^{2}\omega _{p}^{2}\omega _{ce}\Delta \omega _{ce}}{%
16m^{2}c^{4}\left( \omega _{p}^{2}-\Delta \omega _{ce}^{2}\right) }\left[ 1+%
\frac{mv_{th}^{2}}{\hbar \omega _{ce}}\tanh \left( \frac{\mu B_{0}}{%
k_{B}T}\right) \right]  \label{delta-mod}
\end{equation}
We note that $\delta\ll 1$ holds to a very good approximation for most parameters of physical interest.

Next, we consider the spin-resonance mode in the long wavelength regime $%
\left\vert kv_{th}/(\omega -\Delta \omega _{ce})\right\vert \ll 1$, where the
terms proportional to $k^{2}$ of \cref{Dispersion_rel} are small
corrections. The dispersion relation can then be approximated by%
\begin{equation}
\omega =\Delta \omega _{ce}\left[ \left( 1+\delta \right) +\frac{\hbar
^{2}\omega _{p}^{2}}{128m^{2}c^{4}}\frac{\omega _{ce}\Delta \omega _{ce}}{%
\left( \omega _{p}^{2}-\Delta \omega _{ce}^{2}\right) }\frac{k^{2}v_{t}^{2}}{%
\left( \omega -\Delta \omega _{ce}\right) ^{2}}\right]
\label{long-wave spin DR}
\end{equation}

Apparently, since $\delta \ll 1$, and the dispersive term proportional to $k^2$ in \cref{long-wave spin DR} is small in the long wavelength regime, $\omega \approx \Delta \omega_{ce}$ applies for the spin resonance-mode.
Using Eq. (\ref{long-wave spin DR}), we can compute the group velocity for
the spin resonance mode, and compare the magnitude of the classical and the
spin terms in \cref{Ponder_Force}. The first spin term of \cref{Ponder_Force} (proportional to $\partial \left\vert
E\right\vert ^{2}/\partial z$) and  the second spin term (proportional to $%
\partial \left\vert E\right\vert ^{2}/\partial t$) as well as the sum of
both are plotted in \cref{fig3} as a function of normalized wave-number $%
k_{n}=kv_{th}/(\omega -\Delta \omega _{ce})$, where the validity condition
of the plot require $k_{n}\ll 1$. All contributions are normalized against
the classical ponderomotive force, i.e. a contribution equal to $-1$ is
equal in magnitude to the classical ponderomotive force but has the opposite
sign. The classical ponderomotive force for electrostatic fields is always
directed from higher intensity to lower intensity, but, as seen in \cref{fig3},
this does not always hold for the spin contributions. Specifically, the
first spin term, which dominate for the longest wavelength, can have the
opposite sign as the classical term. However, while this term can be significant, 
it cannot be neagtive enough to revert the direction of the total ponderomotive 
force. Thus, independently of wave-number and parameter values, the total 
ponderomotive force for electrostatic fields is always directed from higher to lower intensities.


One might expect that the spin terms are always important for the spin
resonance mode. However, as shown in the first and second panels of \cref{fig3}, 
the classical ponderomotive force can be dominant for the longest wavelengths (both spin terms and the
sum of them are well below unity).  A thing to note when comparing the first and second panels of \cref{fig3} is the change of sign of the first spin term. As it turns out, the first spin term has the same sign as $1-R$. 


Next, as seen in the third panel of \cref{fig3}, we note that the relative importance of the ponderomotive force terms are rather sensitive to the plasma density. Decreasing the density, as captured by the $R$-parameter, we see that the first spin term becomes larger than the classical term, and will dominate in the long wavelength regime.

In \cref{fig3} we have only shown the results for really long wavelengths up to $k_n < 10^{-3}$. As the calculations of this section apply up to $k_n<0.1$, the shorter wavelength regime is also of interest. Using the same parameters as in the second panel of \cref{fig3}, but extending the wave-number regime, we see in the first panel of \cref{fig4} that the spin part of the ponderomotive force (due to the second term) will dominate for the shorter wavelengths. Decreasing the density further (as in the second panel of \cref{fig4} ), the effects are even more pronounced, as the spin term can be more than a factor 50 larger than the classical term. Due to the scale, it is hard to read of the first spin-term that is small compared to the other terms in both panels of \cref{fig4}. This term varies little with wave-numbers and is close to $0.1$ in the upper panel and $0.45$ in the lower panel, for the whole spectrum.


\section{Summary and Conclusions}
In the present paper, we have calculated the ponderomotive force for electrostatic waves propagating in a plasma parallel to an external magnetic field. The calculation has been performed using a quantum kinetic model, including the electron spin dynamics, covering effects such as spin-orbit interaction and Thomas precession. The model is of particular interest for strongly magnetized environments, as can be found in astrophysics. The ponderomotive force is of crucial importance for a large number of nonlinear phenomena, such as e.g. soliton formation, self-focusing, wake field generation, and particle acceleration. 

In section IV wee have studied the relative magnitude of classical and quantum mechanical contributions to the ponderomotive force. An interesting finding is that many of the preliminary conclusions from linear theory does not translate into the nonlinear regime. Thus, even if the inequality \cref{Langmuir inequality} is fulfilled, such that the linear dispersion relation agrees to a good approximation with the classical Langmuir dispersion relation, in the vicinity of the spin resonance the quantum terms may still dominate the expression for the ponderomotive force. Similarly, even when the linear mode is a spin resonance mode (given by expression \cref{long-wave spin DR}, which is quantum mechanical in nature, it may happen that the ponderomotive force is given by the classical expression. However, depending on the plasma parameters and the wave-number, it is also possible that the quantum contribution is larger than the classical one by orders of magnitude. 

Understanding the nonlinear spin dynamics in the simpler case of electrostatic fields is a first step towards understanding more  complex nonlinear phenomena, such as e.g. spin polarization by intense laser pulses \cite{li2019ultrarelativistic,del2017spin}. Moreover, the findings of our paper are a necessary prerequisite for a more detailed analysis of nonlinear phenomena of astrophysical plasmas, in particular accretion discs surrounding objects such as pulsars and magnetars.

 \bibliography{References}

\end{document}